\documentclass[aps,twocolumn,superscriptaddress]{revtex4}
\usepackage{graphicx}

\newcommand{\ket}[1] {\left| #1 \right\rangle}

\begin{document}

\title{Quantum state transfer in a disordered one-dimensional lattice}

\author{S. Ashhab}
\affiliation{Qatar Environment and Energy Research Institute (QEERI), Hamad Bin Khalifa University (HBKU), Qatar Foundation, Doha, Qatar}

\date{\today}


\begin{abstract}
We investigate the effect of disorder on the transfer of quantum states across a one-dimensional lattice with varying levels of control resources. We find that the application of properly designed control signals, even when applied only to the two ends of the lattice, allows perfect state transfer up to disorder strengths that would not allow a generic quantum state to propagate the length of the lattice. At sufficiently large disorder strengths, however, the local control signals fail to send the quantum state from one end of the system to the other end. Our results shed light on the interplay between disorder and controlled transport in one-dimensional systems.
\end{abstract}


\maketitle

\section{Introduction}
\label{Sec:Introduction}

Disorder has a profound effect on transport in quantum systems. One of the most striking manifestations of this statement is Anderson localization \cite{Anderson}, where an increasing amount of disorder can lead to a phase transition from a metallic to an insulating state, even in the seemingly simple case of noninteracting particles. This phenomenon, first predicted over fifty years ago, has been observed directly in a number of physical systems in recent years \cite{AndersonLocalizationExperiments}.

Another subject where there has been extensive work in recent years is the problem of quantum state transfer along one-dimensional lattices or spin chains \cite{Bose,Christandl,BoseReview}. In this problem the goal is to transport an unknown quantum state from one end of a spin chain to the opposite end of the chain. Previous studies have for example considered the probabilistic state transfer in a chain with uniform parameters \cite{Bose}, perfect state transfer with properly designed time-independent system parameters \cite{Christandl}, perfect state transfer using externally applied time-dependent control fields \cite{BurgarthStateTransferWithControl} and quantum state routers in branched spin chains \cite{QuantumRouters}.

One question that has not been investigated in much detail in the literature is the effect of disorder on one's ability to perform perfect state transfer \cite{BurgarthStateTransferWithDisorder1,BurgarthStateTransferWithDisorder2,Wu}. One can say that there are some simple limiting cases and trickier, intermediate cases in relation to this question. If, for example, one has full control on single-spin parameters but there is disorder in the inter-spin coupling strengths, one is able to perform perfect state transfer by using the large amount of control resources to counter the adverse effects of the disorder (e.g.~by performing a sequence of two-spin swap operations while decoupling the rest of the chain during each such two-spin operation). Even in this case, where intuition tells us that perfect state transfer must be possible (and does not involve exponential scaling of resources), there remains the practically important question of the minimum time required for perfect state transfer. On an opposite extreme, if one uses a protocol that relies on static Hamiltonian settings, e.g.~as is done in Ref.~\cite{Christandl}, then one would intuitively infer that disorder will induce Anderson localization and state transfer will not be possible for long chains. An intermediate case is that where one has full control over the parameters at the two opposite ends of the chain, with disorder present throughout the chain. In this case there are spatial limitations on the control resources, but there are no temporal limitations. The question then arises whether this high degree of control in the time domain will for example allow one to properly launch a spin wave that carries the quantum state and is able to propagate through the disordered landscape of the spin chain without suffering Anderson localization. In order to appreciate this question, one could note here that the problem of Anderson localization is usually studied in a control-free setting where the wave packet is assumed to be a generic one by some definition and in particular not prepared for the specific disorder landscape that it is going to encounter. As a result, it is difficult to take intuition accumulated from past studies on Anderson localization and use it to predict the effect of disorder on systems that are subjected to driving by control signals.

In this paper we present results of numerical simulations on the propagation and transfer of quantum states through a disordered chain with and without control. In the case of controlled state transfer, we use optimal control theory to determine the control signals that maximize the transfer fidelity. We find that the control signals can help suppress the effects of small amounts of disorder and enable fast perfect state transfer up to a reasonably large disorder strength, even with a limited amount of control. For large disorder strengths, however, Anderson localization physics eventually sets in and the controlled transfer cannot be improved beyond the sequential nearest-neighbor transfer protocol, and in the case of limited control resources the transfer time grows indefinitely. For intermediate values of the disorder strength, we find large variations in the transfer fidelity for different disorder instances, indicating that different disorder patterns can have drastically different localizing effects, even for the same disorder strength.

The remainder of this paper is organized as follows: In Sec.~\ref{Sec:Setup} we describe the model system and formulate the problem to be solved. In Sec.~\ref{Sec:Results} we present the results of our numerical simulations and discuss their implications. Section \ref{Sec:Conclusion} contains concluding remarks.

\section{Problem setup}
\label{Sec:Setup}

\begin{figure}[h]
\includegraphics[width=8.5cm]{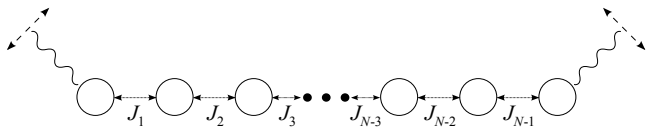}
\caption{Schematic diagram of a chain with $N$ spins, nearest-neighbor inter-spin coupling strengths $J_i$ and control fields applied to the first and last spins in the chain. In this system, the on-site energies are equal throughout the lattice, except for the edge spins, where the on-site energies $\omega_1$ and $\omega_N$ can be tuned via the control fields.}
\label{Fig:SchematicDiagram}
\end{figure}

We consider a spin chain described by the Hamiltonian:
\begin{equation}
\hat{H} = \sum_{i=1}^{N} \frac{\omega_i}{2} \hat{\sigma}_z^{(i)} + \sum_{i=1}^{N-1} J_i \hat{\sigma}_x^{(i)} \otimes \hat{\sigma}_x^{(i+1)},
\label{Eq:SpinChainHamiltonian}
\end{equation}
where $\omega_i$ are the on-site energies, $J_i$ are the (nearest-neighbor) inter-spin coupling strengths and $\sigma_{\alpha}^{(i)}$ (with $\alpha=x,y$ or $z$) are the usual Pauli operators for the spin at site $i$. Disorder can be introduced into the problem through $\omega_i$ and/or $J_i$, and Anderson localization occurs in both cases, up to minor quantitative differences. We shall assume that the control fields couple to single-spin operators, in particular assuming that some or all $\omega_i$ are tunable parameters. We therefore focus on the case where disorder enters the problem through $J_i$. We use a uniform distribution for $J_i$, i.e.~for a disorder strength $\Delta$ the values of $J_i$ are distributed uniformly between $\overline{J}(1-\Delta)$ and $\overline{J}(1+\Delta)$. As our main setup for discussions, we shall assume that the on-site energies at the beginning and end of the chain, i.e.~$\omega_1$ and $\omega_N$, are control parameters that can be adjusted without any constraints, while all other $\omega_i$ are fixed and uniform, i.e.~$\omega_2=\omega_3=...=\omega_{N-1}$. The system is illustrated in Fig.~\ref{Fig:SchematicDiagram}.

The objective of the control protocol is to transfer a quantum state from the first site to the last site of the chain. The chain is therefore assumed to initially be in the state
\begin{equation}
\ket{\Psi_{\rm initial}} = \ket{\phi}_1 \otimes \ket{\rm ground \ state}_{2,3,\dots,N},
\label{Eq:PsiInitial}
\end{equation}
and the desired final state is
\begin{equation}
\ket{\Psi_{\rm target}} = \ket{\rm arbitrary \ state}_{1,2,\dots,N-1} \otimes \ket{\phi}_N,
\label{Eq:PsiFinal}
\end{equation}
where the indices label the different sites in the chain. The ground state is used in Eq.~(\ref{Eq:PsiInitial}) because it is typically easy to prepare. The state $\ket{\phi}$, which can be expressed as $\alpha\ket{0}+\beta\ket{1}$, is treated as an unknown quantum superposition of the ground and excited states. There are a number of different scenarios where the assumption of an unknown state arises. For example, the sender could have received the state $\ket{\phi}$ as a bit of quantum information from another source or obtained it as the output of a quantum information protocol, and the sender is therefore unable to know the quantum state without disturbing it. Note that the state could be entangled with other parts of the system, and a protocol that transforms Eq.~(\ref{Eq:PsiInitial}) into Eq.~(\ref{Eq:PsiFinal}) will similarly transfer any entanglement associated with the quantum state. The state could also be known to the sender, e.g.~the state $\ket{\uparrow}$ or $\ket{\downarrow}$ along an axis that the sender chooses but does not reveal to anybody else until a later point in the protocol, and in order to protect this information against eavesdropping any part of the state transfer protocol outside of the sender's location must be designed to work for an arbitrary quantum state.

If we start with an unknown quantum state at the first site with the rest of the chain prepared in its ground state, i.e.~as in Eq.~(\ref{Eq:PsiInitial}), the subsequent dynamics will in general involve complex dynamics in the $2^N$-dimensional Hilbert space of the $N$-site chain. Since this problem becomes intractable for chains of length $N \sim 20$, we assume, as is commonly done in this context, that the parameters are in the regime where the rotating-wave approximation is valid. In other words, we assume that $J_i$ are much smaller than $\omega_i$, such that to a good approximation we can ignore terms in the Hamiltonian given in Eq.~(\ref{Eq:SpinChainHamiltonian}) that mix states with different values of $\sum_{i=1}^N\sigma_z^{(i)}$, and we can use the approximate Hamiltonian 
\begin{equation}
\hat{H} = \sum_{i=1}^{N} \frac{\omega_i}{2} \hat{\sigma}_z^{(i)} + \sum_{i=1}^{N-1} J_i \left[ \hat{\sigma}_+^{(i)} \otimes \hat{\sigma}_-^{(i+1)} + \hat{\sigma}_-^{(i)} \otimes \hat{\sigma}_+^{(i+1)}\right],
\label{Eq:SpinChainHamiltonianRWA}
\end{equation}
where $\sigma_{\pm}^{(i)}$ are the spin raising and lowering operators at site $i$. Under this approximation, the number of excitations in the system becomes a conserved quantity. As a result, and since the global ground state $\ket{0}_1 \otimes \ket{0}_2 \otimes \cdots \otimes \ket{0}_{N-1} \otimes \ket{0}_N$ (with $\ket{0}_i$ being the ground state of the single-site Hamiltonian at site $i$) does not evolve except for acquiring a simple phase, the problem of transferring an unknown single-spin quantum state from the first to the last spin (with all other spins starting and ending up in their ground states) can be simplified to the problem of transferring a single excitation from the first to the last site in the chain, i.e.~transforming the state $\ket{\psi}=\ket{1}\equiv \ket{1}_1 \otimes \ket{0}_2 \otimes \cdots \otimes \ket{0}_{N-1} \otimes \ket{0}_N$ into the state $\ket{\psi}=\ket{N}\equiv \ket{0}_1 \otimes \ket{0}_2 \otimes \cdots \otimes \ket{0}_{N-1} \otimes \ket{1}_N$. The excitation that is transferred can alternatively be thought of as a particle. In other words, one has a natural mapping between the spin chain containing a single excitation and a system containing a single particle hopping between lattice sites. In both cases the system dynamics is well described using the picture of wave propagation. The Hilbert space required in order to study these problems is of size $N$ only. The simplification of the problem from a $2^N$-dimensional to an $N$-dimensional Hilbert space allows us to investigate relatively long chains, well above the limit of $N \sim 20$ when dealing with $2^N$-dimensional Hilbert spaces.

Since the output of the protocol is the state at site $N$, we in principle evaluate the performance of the protocol based on how close the state of site $N$ at the final time (after tracing out the state of all the other sites) is to the state $\ket{\phi}$, averaged over all possible single-spin states $\ket{\phi}$. As mentioned above, however, under the rotating-wave approximation the transfer fidelity can also be evaluated through the probability $P$ that the initial state $\ket{\psi}=\ket{1}$ evolves to the state $\ket{\psi}=\ket{N}$ at the final time. This probability can be expressed as
\begin{equation}
P = | \langle \psi(t_{\rm final}) \ket{N} |^2.
\end{equation}

As mentioned above, in our main simulations we shall assume that $\omega_1$ and $\omega_N$ are tunable parameters, with all the other $\omega_i$ fixed. In this case $\omega_1$ and $\omega_N$ can be treated as control parameters, and we can use optimal control theory techniques to identify the control signals that maximize the transfer fidelity. We use the GRAPE algorithm \cite{Khaneja} to obtain the optimal control pulses \cite{PulseFootnote}. In each such calculation, we search for the control pulse that maximizes the probability $P$. The idea is to perform these calculations for a few different values of the pulse time. The probability $P$ increases from zero at very short pulse times to values that are essentially equal to unity for pulse times that are equal to or larger than the minimum perfect-transfer time \cite{Ashhab}. We shall refer to this time as $T_{\rm min}$.

In all of the calculations presented here, we consider chains containing fifty sites. We divide the total pulse time $T$ into 200 time steps \cite{GRAPEFootnote}. We perform the GRAPE algorithm with up to 4000 iterations, although in most cases the fidelity saturated and stopped increasing within the first few hundred iterations \cite{ComputationTimeFootnote}. It should be noted that each pulse that we find using this technique is optimized for the particular set of parameters in a given instance of disorder and will not be optimal for other instances with the same disorder strength. These instance-specific optimized pulses are relevant in cases where the disorder is static, e.g.~disorder related to defects that appear during the fabrication of a device but do not change in time. It is crucial to note in this context that the coupling strengths in the chain can be measured using access to only the ends of the chain \cite{BurgarthCouplingStrengthEstimation}, such that it is not unrealistic to assume that one has access to only the edges of the chain but is nevertheless able to have knowledge about the values of $J_i$ for the entire chain. It is also crucial to note that the optimized pulses are independent of the quantum state that is transferred across the chain.

A natural reference point that is useful to set at this point is the time that it takes to transfer a state from one end of the chain to the opposite end using a sequence of swap operations, one step at a time. Each such swap operation takes $\pi/(2J_i)$, such that the transfer across the entire chain of length $N$ takes a total time of
\begin{equation}
T_{\rm seq}=\sum_{i=1}^{N-1}\frac{\pi}{2J_i} = \frac{(N-1)\pi}{2}\times\overline{J_i^{-1}}.
\end{equation}
In the absence of disorder and with full control over all $\omega_i$, one can perform the state transfer more quickly by tuning the parameters in such a way that one launches a spin wave that traverses the chain at the maximum possible wave speed (keeping in mind that in such discrete systems there is a maximum propagation speed set by $J$), leading to the result that $T_{\rm min}=T_{\rm seq}/\pi$ in the limit of an infinitely long chain (see e.g.~Ref.~\cite{Ashhab}).

\section{Results and discussion}
\label{Sec:Results}

As mentioned in Sec.~\ref{Sec:Setup}, the probability $P$ of successful transfer of a single excitation across the chain is a good measure of the transfer fidelity for an unknown quantum state, and we shall use this probability in our analysis below. More specifically, we assume that at the initial time an excitation is localized at the first site in the chain, and we investigate the probability that the excitation will be found at the last site in the chain at the end of a given pulse time.

\begin{figure}[h]
\includegraphics[width=8.0cm]{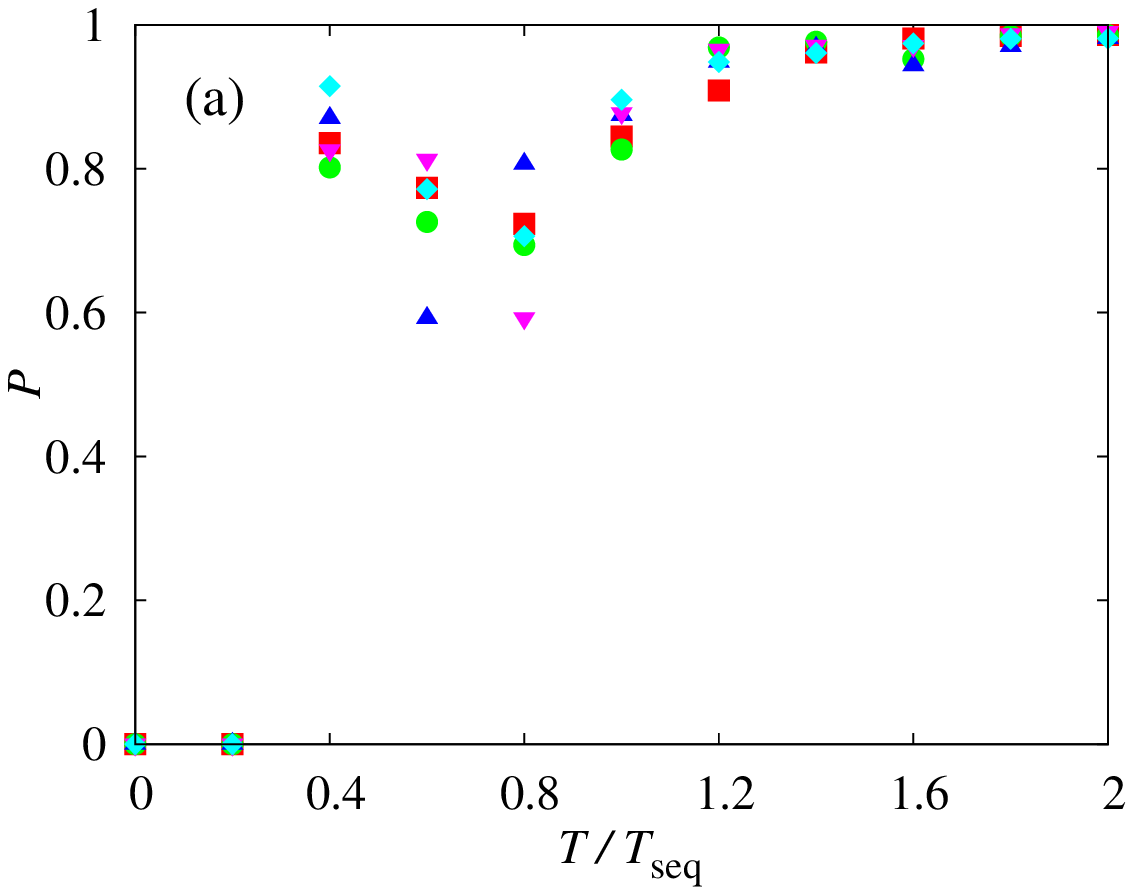}
\includegraphics[width=8.0cm]{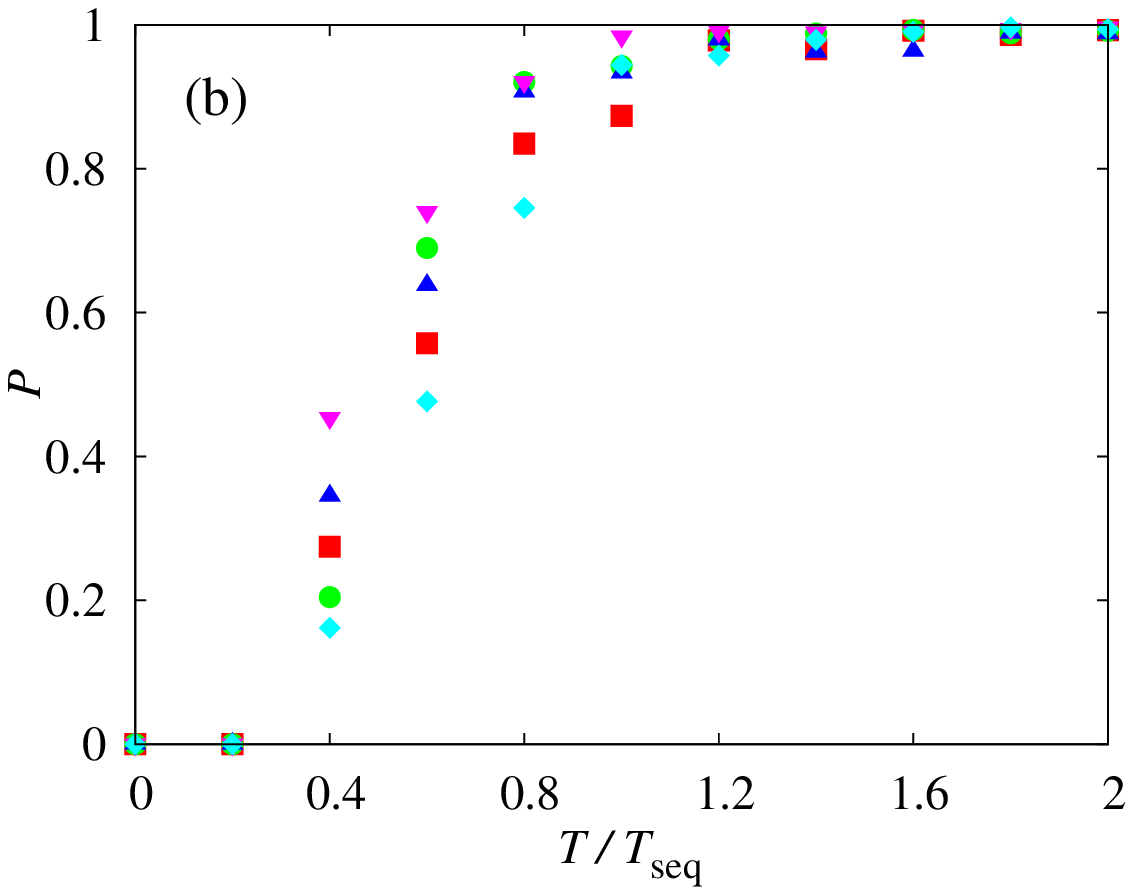}
\includegraphics[width=8.0cm]{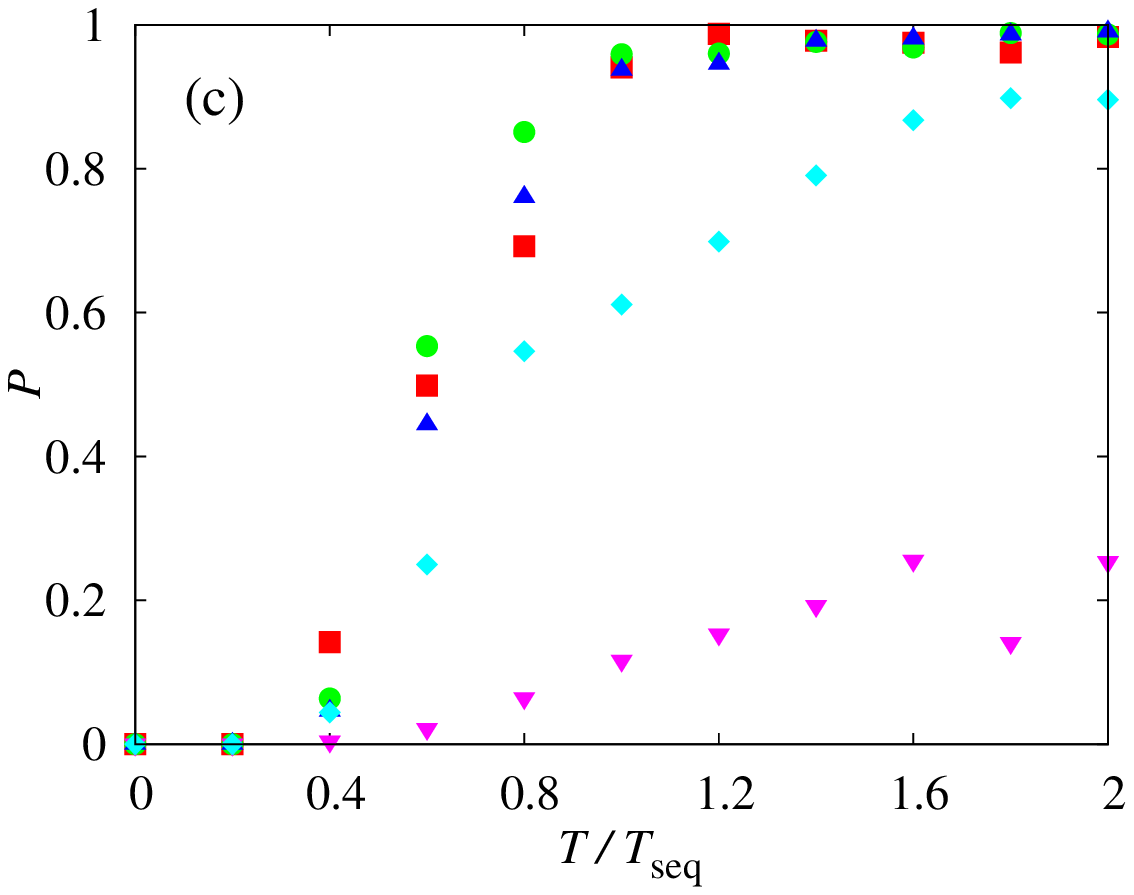}
\caption{The transfer probability $P$ as a function of pulse time $T$ for a fifty-site chain and control over the two edge sites. The time is measured relative to the sequential swap transfer time $T_{\rm seq}$. The different panels correspond to different values of the disorder strength: $\Delta=0.1$ (a), 0.3 (b) and 0.5 (c). There are five sets of data points in each panel. These correspond to five different instances of the disordered coupling strengths that we generate for each value of coupling strength.}
\label{Fig:ProbabilityAsFunctionOfTime}
\end{figure}

In Fig.~\ref{Fig:ProbabilityAsFunctionOfTime} we plot $P$ as a function of pulse time $T$ for a few different instances of disorder with varying disorder strength in the case where control fields are applied to only the first and last sites in the chain. As would be expected, $P$ increases from zero at small values of $T$ to unity at a value that defines the minimum perfect-transfer time $T_{\rm min}$, and it remains equal to unity for larger values of $T$ \cite{NonmonotonicProbabilityFootnote}. For $\Delta=0.1$, where the disorder is weak, $P$ jumps from almost zero at pulse time $T=0.2\times T_{\rm seq}$ to almost unity at $T=0.4\times T_{\rm seq}$, which means that $T_{\rm min}$ is close to $T_{\rm seq}/\pi$, the limiting value in the case of vanishing disorder strength and full control over the on-site energies. Even at $\Delta=0.3$, the disorder does not seem to cause much slowing down of the transfer and the different instances of disorder do not lead to large variations in $T_{\rm min}$. Only one out of five disorder instances in this case (magenta triangles) seems to give a transfer probability that is consistently higher than those in the other four instances, and the difference is rather small. It should be noted, however, that $T_{\rm min}$ is now close to $T_{\rm seq}$, which is a factor of $\pi$ slower than that obtained in the absence of disorder. A more dramatic change occurs when we increase the disorder strength $\Delta$ to 0.5. In three out of five disorder instances, $T_{\rm min}$ is still around $T_{\rm seq}$, but in one instance there is a slowing down by a factor of two and in another instance the transfer seems to slow down by at least an order of magnitude. These large variations can be intuitively understood by keeping in mind that some instances of the disorder will contain potential barrier landscapes that are more localizing than others.

\begin{figure}[h]
\includegraphics[width=8.5cm]{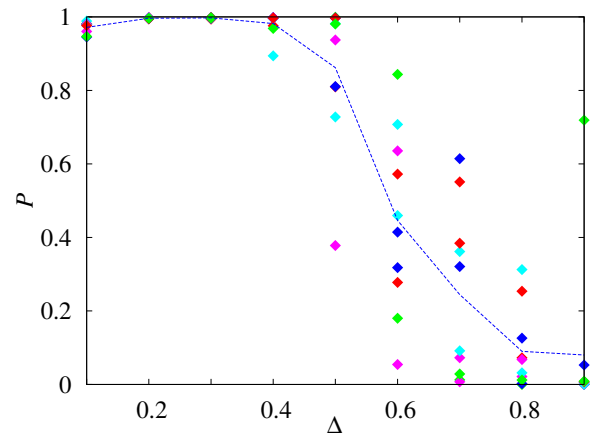}
\caption{The transfer probability $P$ as a function of disorder strength $\Delta$ for a fifty-site chain, a pulse time $T=T_{\rm seq}=(N-1)\pi/2\times\overline{1/J_i}$ and control over the two edge sites. The ten different data points for each value of $\Delta$ correspond to ten different instances of the randomly generated coupling strengths. The different colors are used in order to help resolve closely spaced points but have no physical significance. The dashed line shows the mean value of $P$ averaged over the ten different instances at each value of $\Delta$.}
\label{Fig:ProbabilityAsFunctionOfDisorderStrength}
\end{figure}

From Fig.~\ref{Fig:ProbabilityAsFunctionOfTime} we can see that starting from $T=T_{\rm seq}/\pi$ the transfer probability $P$ increases steadily until it reaches a value very close to unity. Furthermore, in instances where the disorder is not causing a serious increase in $T_{\rm min}$, $P$ has values close to unity at $T=T_{\rm seq}$. We therefore perform further calculations focusing on this value of $T$ as a way to get a simple but rather reliable indicator for $T_{\rm min}$. In Fig.~\ref{Fig:ProbabilityAsFunctionOfDisorderStrength} we plot $P$ as a function of disorder strength $\Delta$ for $T=T_{\rm seq}$. Up to $\Delta=0.4$, $P$ remains high for the vast majority of disorder instances. Between $\Delta=0.5$ and $\Delta=0.7$, different instances of disorder lead to drastically different values of $P$. Above $\Delta=0.7$, $P$ quickly becomes small for the vast majority of disorder instances (with one notable exception where $\Delta=0.9$ and $P=0.71$). As we shall see below, the disorder strength range where the probability drops from high to low (i.e.~between 0.5 and 0.7, roughly speaking) is related to the range where the Anderson localization length becomes much smaller than the chain length.

The above results indicate that the spatially constrained but temporally unconstrained control that is applied to the two edges of the chain is successful in suppressing the localizing effects of disorder for small values of the disorder strength, but it eventually fails for strong disorder. A possible intuitive explanation for this phenomenon is that for weak disorder a wave packet (or at least part of it) is able to propagate the full length of the chain, and the control fields are then able to ensure that the full amplitude of the wave packet is collected at the last site in the chain at the end of the pulse (with a pulse time that grows only linearly with chain length). For strong disorder, the wave packet cannot reach the end of the chain, and at least as far as the last site of the chain is concerned there is nothing that the control fields can do in order to catch the wave packet there.

In order to identify more clearly the role of the control pulses in countering the localizing effects of disorder, we perform a few further sets of calculations with different assumptions regarding the control resources. First we consider the problem of state transfer without control or, in other words, the free propagation of waves. We also analyze the spatial extension of energy eigenstates in the presence of disorder. We then consider the case where one has control over a number of sites located equidistantly from each other (up to the minor constraint imposed by the exact chain length) and the case where one has full control over all the sites in the chain.

\begin{figure}[h]
\includegraphics[width=6.5cm]{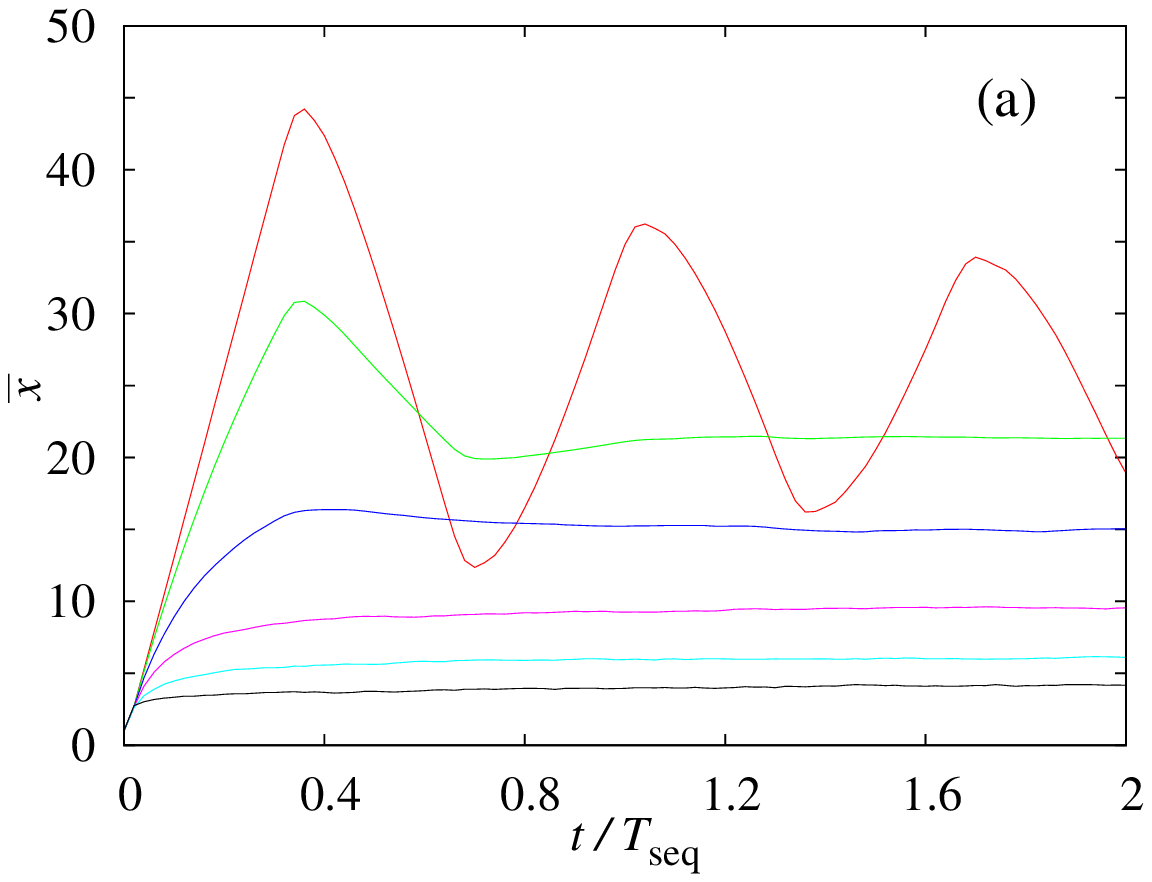}
\includegraphics[width=7.0cm]{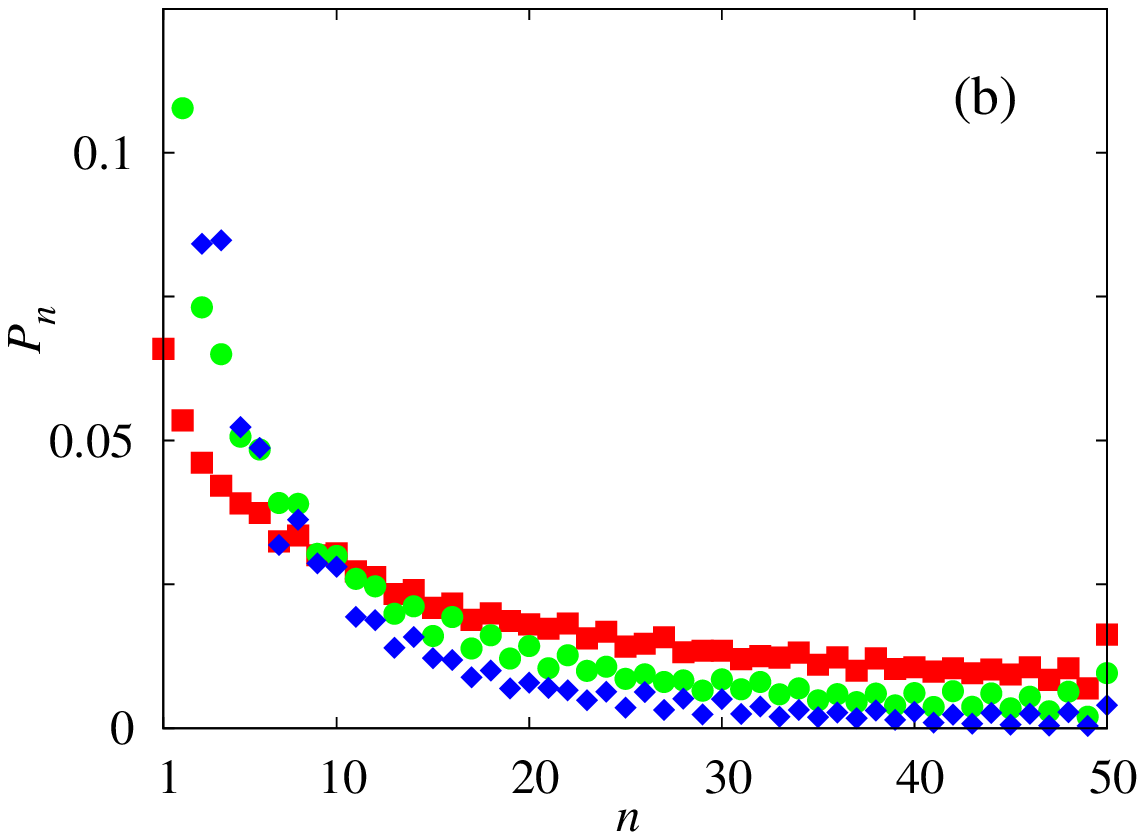}
\includegraphics[width=7.0cm]{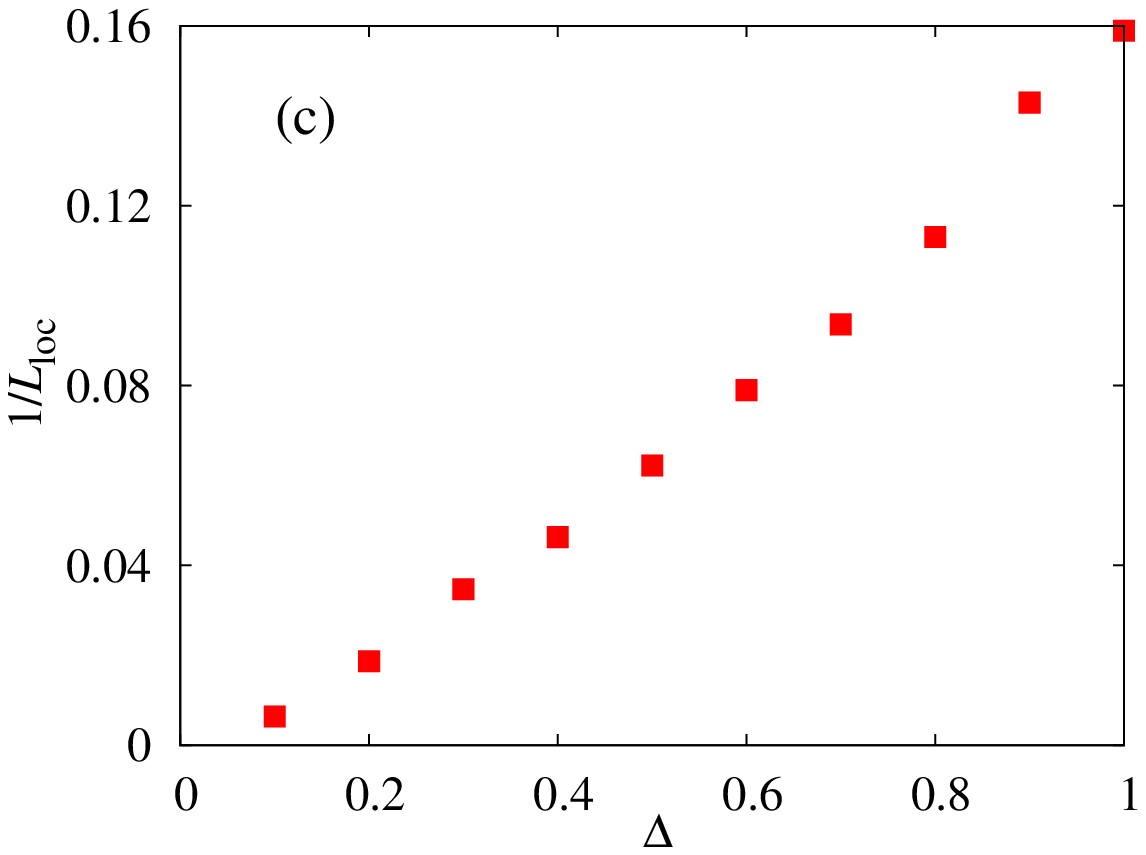}
\caption{(a) Average position $\overline{x}$ as a function of time $t$ for an excitation that is initialized at site 1 and propagates freely in a fifty-site chain, averaged over $10^3$ disorder instances. The different curves correspond to different values of disorder strength: from top to bottom, $\Delta=0,0.2,0.4,...,1$. (b) The long-time occupation probability $P_n=\left|\psi_n(2T_{\rm seq})\right|^2$ as a function of site index $n$, showing the exponential-decay behavior. The time $t=2T_{\rm seq}$ is chosen because, as can be seen from panel a, it is sufficiently long that a steady-state distribution has been reached. The three different symbols correspond to different values of disorder strength: $\Delta=0.2$ (red squares), 0.4 (green circles) and 0.6 (blue diamonds). (c) The inverse of the localization length $L_{\rm loc}$ that is extracted from the exponentially decaying long-time probability distribution as a function of disorder strength $\Delta$.}
\label{Fig:FreePropagation}
\end{figure}

In the first set of these additional calculations, we initialize the system in the state $\ket{\psi}=\ket{1}$ with an excitation (or particle) localized at the first site, and we simply let it evolve according to the Schr\"odinger equation with the Hamiltonian given in Eq.~(\ref{Eq:SpinChainHamiltonianRWA}) and time-independent parameters. For a given value of $\Delta$, we run $10^3$ different instances of the disordered Hamiltonian and inspect physical observables averaged over these instances. In particular, we inspect the average position as a function of time $\left[\overline{x}(t)=\sum_{n}n\times\left|\psi_n(t)\right|^2\right]$ and the site occupation probabilities as functions of position and time. The results are plotted in Fig.~\ref{Fig:FreePropagation}. The average position $\overline{x}(t)$ clearly shows that the excitation propagates and possibly bounces back and forth between the ends of the chain before it converges to a constant value. This long-time value decreases with increasing disorder strength, which suggests that it is related to the Anderson localization length. Furthermore, after the transient propagation period (whose duration also shrinks with increasing disorder strength), the (averaged) probability distribution stabilizes and follows an exponential decay function as a function of site index (with only the first and last sites remaining slightly above the exponential fitting function). From this long-time probability distribution we can extract a localization length $L_{\rm loc}$. Figure \ref{Fig:FreePropagation}(c) shows a plot of $1/L_{\rm loc}$ as a function of $\Delta$. One can see that $1/L_{\rm loc}$ increases from a value that is on the order of $1/N$ for small amounts of disorder to a value close to 1/6 for $\Delta=1$. This latter value indicates that the excitation propagates at most a few sites away from its initial location, hence clearly showing Anderson localization. The reason why we obtain a finite propagation distance even for the maximum value of the disorder strength in our calculations, i.e.~$\Delta=1$, is the fact that even for this maximum value of $\Delta$ there is a reasonable probability to have a smooth landscapes up to a certain distance from the chain edge, such that the excitation propagates a few sites on average. Apart from the overall scale, the exact value of $1/L_{\rm loc}$ must be specific to the choice of the distribution function for $J_i$ values. It should be noted here that disorder in $\omega_i$ would not have an upper limit, in contrast to the case with disorder in $J_i$, and one can expect that $1/L_{\rm loc}$ will grow indefinitely with increasing disorder in that case.

Using the results for the localization length in Fig.~\ref{Fig:FreePropagation}, we can evaluate that the mean value of $P$ in Fig.~\ref{Fig:ProbabilityAsFunctionOfDisorderStrength} has the value 0.5 when $L_{\rm loc}=13\approx N/3.8$. Similar calculations (whose results are not shown here in detail) with $N=30$ and 40 give similar values for the ratio $N/L_{\rm loc}$ (3.1 and 3.6, respectively), which gives further support to the idea that the drop of $P$ from near unity to near zero is correlated with the range where the localization length becomes significantly smaller than the chain length.

A point that is worth noting here is that in the strong-disorder regime $\overline{x}$ reaches a steady-state value that is much smaller than $N$, and here $N=50$. As such,the diffusive transport regime (where $\overline{x}$ increases as $\sqrt{t}$) is not realized. If we include decoherence or time-dependent fluctuations, we would obtain such diffusive dynamics. In the weak-disorder regime, we have ballistic transport (with $\overline{x}\propto t$) during the initial transient period, as can be seen in Fig.~\ref{Fig:FreePropagation}(a).

\begin{figure}[h]
\includegraphics[width=9.5cm]{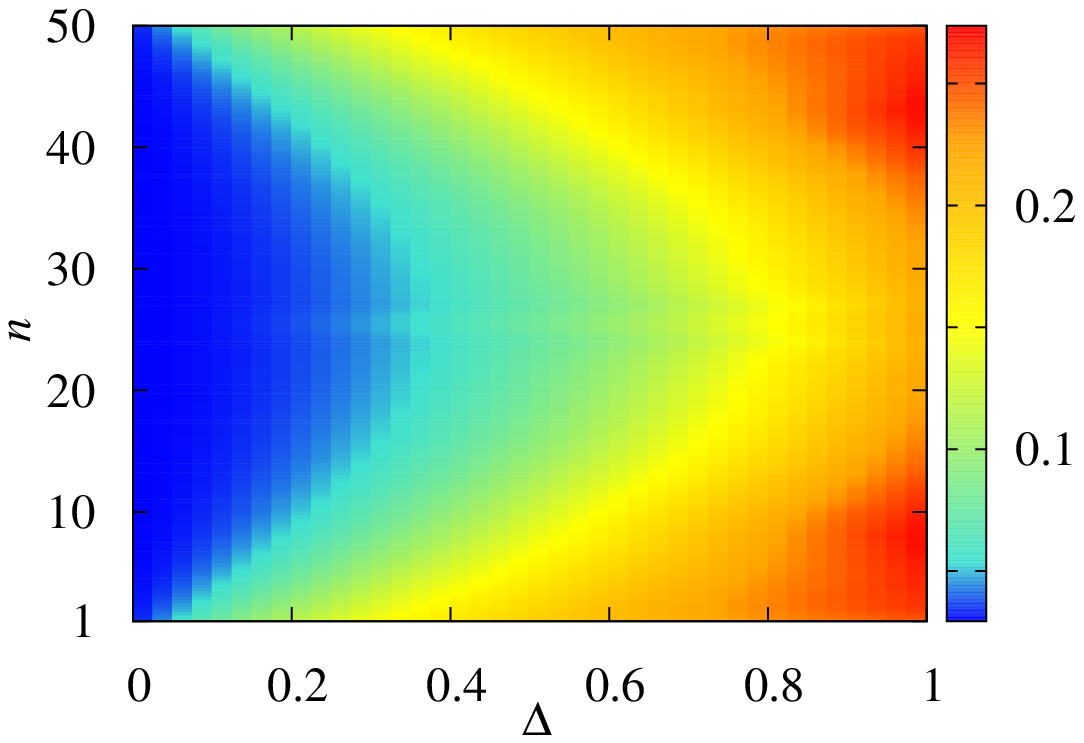}
\includegraphics[width=8.5cm]{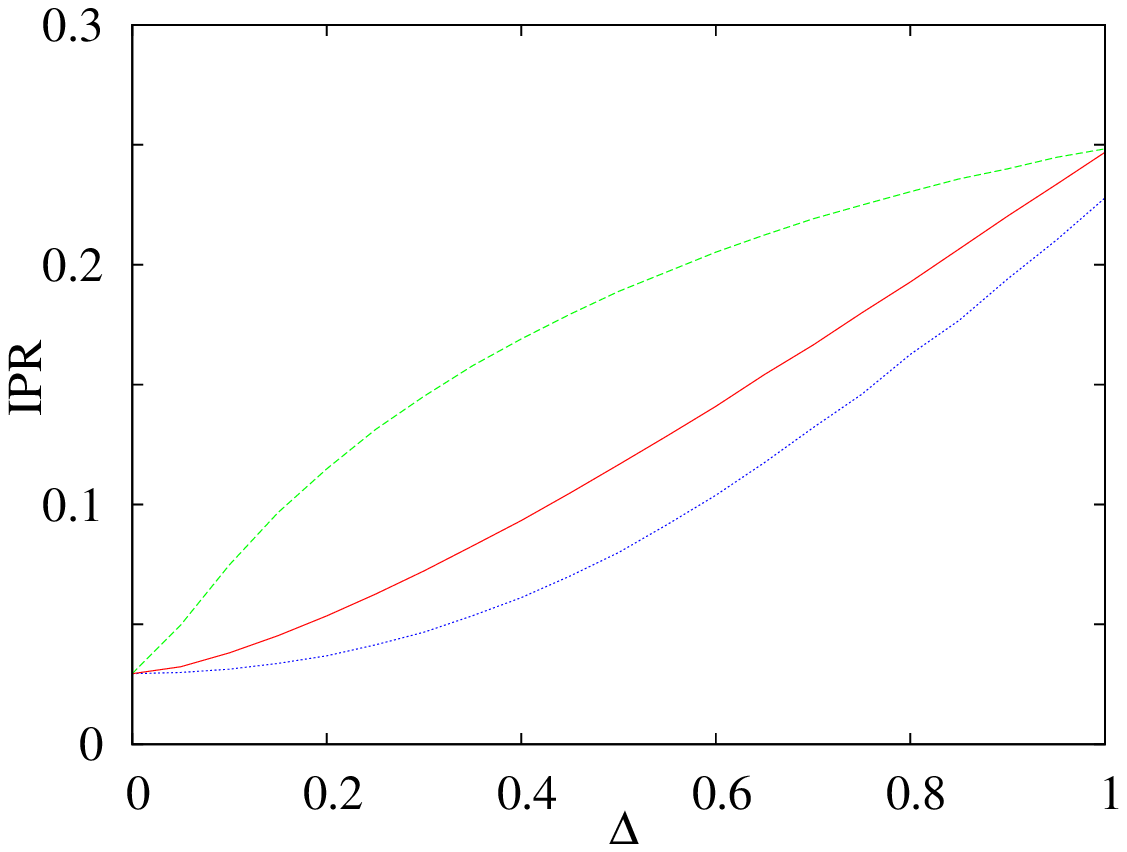}
\caption{Upper panel: The inverse participation ratio (IPR) as a function of disorder strength $\Delta$ and energy level index $n$. Lower panel: The IPR as a function of $\Delta$ for two representative energy eigenstates, namely $n=1$ (green dashed line) and $n=20$ (blue dotted line), as well as the average IPR for the fifty states (red solid line). One can see that all energy eigenstates progress from being delocalized in the absence of disorder (with the IPR being comparable to $1/N$) to being localized with a spatial extent of only a few lattice sites when $\Delta=1$. The states near the edges of the spectrum are more susceptible to disorder-induced localization than states in the middle of the spectrum.}
\label{Fig:InverseParticipationRatio}
\end{figure}

In addition to the localization obtained in the free propagation of initially localized states, localization effects also manifest themselves in the energy eigenstates. These effects can be investigated through the inverse participation ratio (IPR), which for a given quantum state $\psi$ is given by:
\begin{equation}
{\rm IPR} = \frac{\sum_{n=1}^N\left|\psi_n\right|^4}{\left(\sum_{n=1}^N\left|\psi_n\right|^2\right)^2},
\end{equation}
where the index $n$ runs over all the sites in the chain. Roughly speaking, the IPR for a state gives the inverse of the spatial extent of the state. Note that there are $N$ different values of the IPR for the $N$ energy eigenstates of the system \cite{IPR}. The instance-averaged IPR as a function of disorder strength and energy level index is plotted in Fig.~\ref{Fig:InverseParticipationRatio}. The increase in the IPR mirrors that in $1/L_{\rm loc}$, up to a factor of order unity, which shows that Anderson localization manifests itself similarly in these two problems.

\begin{figure}[h]
\includegraphics[width=8.5cm]{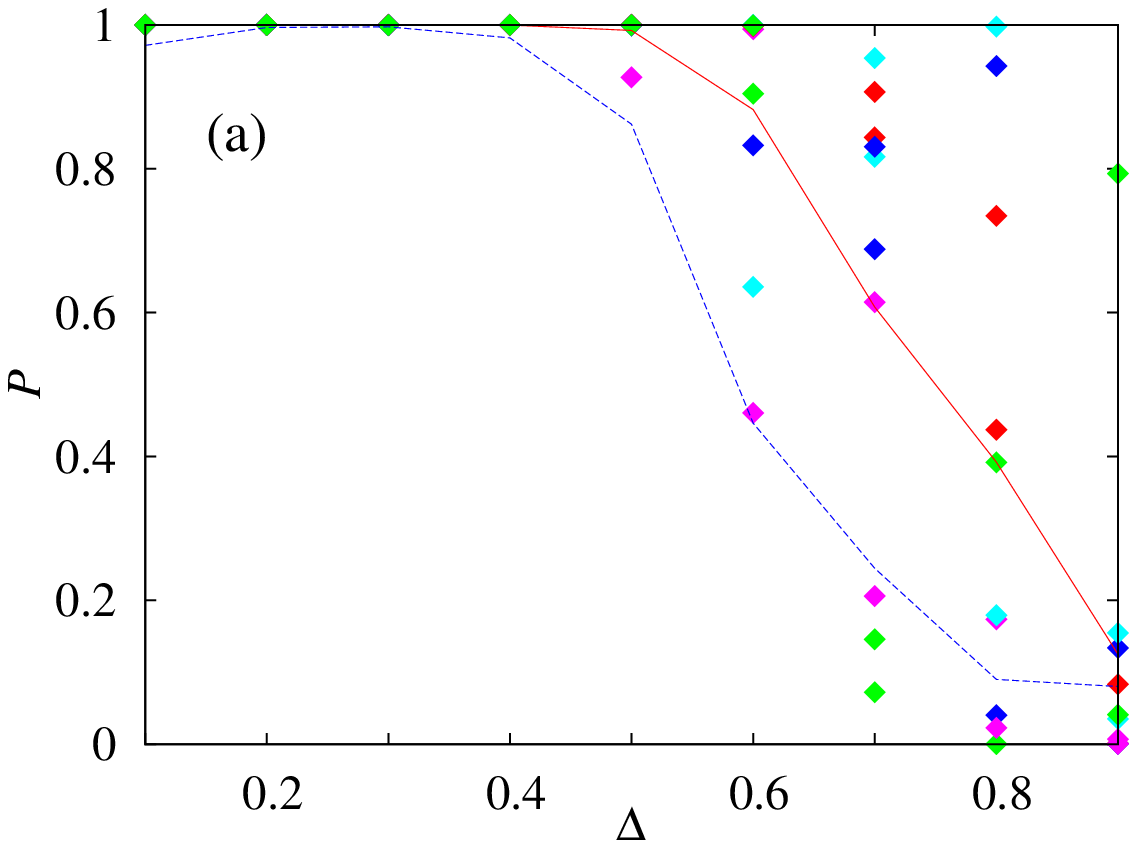}
\includegraphics[width=8.5cm]{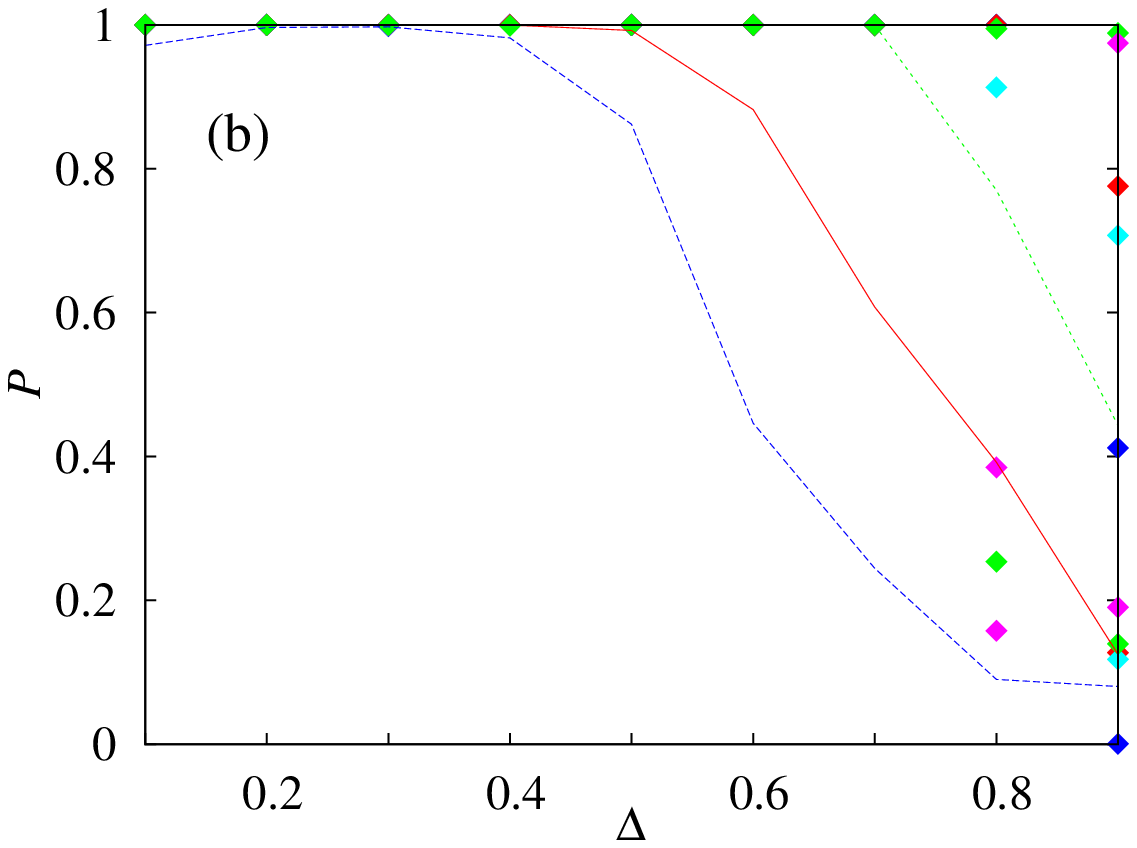}
\caption{The transfer probability $P$ as a function of disorder strength $\Delta$ for a fifty-site chain, a pulse time $T=T_{\rm seq}$ and control over four sites (namely sites 1, 18, 35 and 50; panel a) or eight sites (namely sites 1, 8, 15, 22, 29, 36, 43 and 50; panel b). The ten different data points for each value of $\Delta$ correspond to the same ten instances of the randomly generated coupling strengths used in Fig.~\ref{Fig:ProbabilityAsFunctionOfDisorderStrength}. The different colors are used in order to help resolve closely spaced points but have no physical significance. The solid red line shows the mean value of $P$ for the four-site-control protocol averaged over the ten different disorder instances at each value of $\Delta$. The dotted green line shows the mean value of $P$ for the eight-site-control protocol. The dashed blue line shows the mean value of $P$ for the case where one has control over the edge sites only, i.e.~the same as in Fig.~\ref{Fig:ProbabilityAsFunctionOfDisorderStrength}.}
\label{Fig:ProbabilityAsFunctionOfDisorderStrengthWithIntermediateSiteControl}
\end{figure}

Next we consider the case where control fields are applied to intermediate sites located equidistantly in the chain as well as the two edge sites (i.e.~sites 1 and 50). This situation is somewhat similar to that encountered in the context of quantum repeaters, where one has multiple communication stations located at properly chosen locations between two communicating parties, such that the distance from each station to the next one is sufficiently small that quantum information can be exchanged efficiently between the two parties located at opposite ends of the setup. The results of our calculations are plotted in Fig.~\ref{Fig:ProbabilityAsFunctionOfDisorderStrengthWithIntermediateSiteControl}. In the first set of calculations, we consider the situation where control fields are applied to two intermediate sites in addition to the edge sites. The largest gains obtained with the enhanced control resources are seen between $\Delta=0.5$ and $\Delta=0.8$, which is the range where the transfer probability with only edge control drops from near unity to near zero. The gains start shrinking for stronger disorder, however, and for $\Delta=0.9$ one hardly gains any increase in the transfer probability. When we increase the number of controlled sites from four to eight, meaning that the distance between each controlled site and the next one is only seven sites, the improvement is enhanced further. The transfer probability is now very high in all instances up to $\Delta=0.7$, and even at $\Delta=0.8$ the transfer probability is higher than 0.9 in the majority of the instances. Although we now see a clear enhancement at $\Delta=0.9$ compared to the cases with fewer controlled sites, with a fidelity above 95\% in two of the ten instances, the transfer probability still shows a clear tendency to drop towards zero as we approach $\Delta=1$.

\begin{figure}[h]
\includegraphics[width=8.5cm]{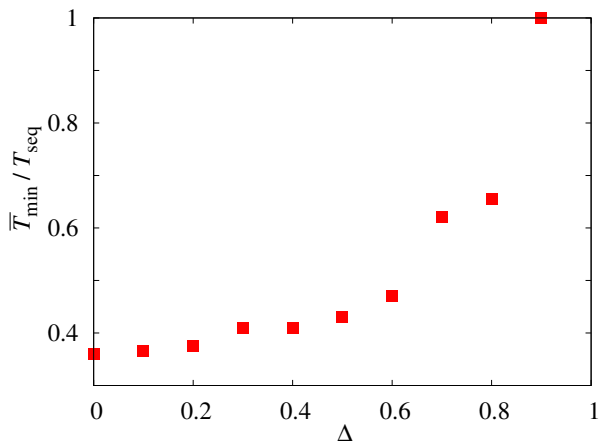}
\caption{The average value of the minimum transfer time $\overline{T}_{\rm min}$ as a function of disorder strength $\Delta$ for the case where all the on-site energies are tunable and treated as control variables. Each point in the figure is a plot over five disorder instances used in the calculations for Fig.~\ref{Fig:ProbabilityAsFunctionOfDisorderStrength}.}
\label{Fig:AverageTransferTimeWithFullOnsiteControl}
\end{figure}

Finally we perform calculations where we determine the minimum perfect-transfer time $T_{\rm min}$ in the case where control fields are applied to all the on-site energies in the chain. As mentioned above, in the absence of disorder and in the limit of an infinitely long chain, $T_{\rm min}=T_{\rm seq}/\pi$. Our numerical calculations with $N=50$ give $T_{\rm min}=1.15\times T_{\rm seq}/\pi$. The 15\% difference with the theoretical formula can be attributed to a combination of finite-size effects (i.e.~the finite length of the chain), a slightly early termination of the search algorithm (noting here that the algorithm exhibits slow convergence when it approaches the maximum transfer probability) and the finite distance between pulse-time values used in our calculations (which is about 4\%). We performed similar calculations for finite values of disorder strength and calculated the average value of $T_{\rm min}$ (averaged over a few disorder instances), to which we refer as $\overline{T}_{\rm min}$. The results are shown in Fig.~\ref{Fig:AverageTransferTimeWithFullOnsiteControl}. The full control over the on-site energies allows one to strongly suppress the localizing effects of the disorder in the coupling strengths up to $\Delta=0.6$, after which $\overline{T}_{\rm min}$ increases rapidly. Note here that the sequential swap protocol gives a transfer time $T_{\rm seq}$, providing an upper bound for $\overline{T}_{\rm min}$. In other words, even for very strong disorder, one can still perform perfect state transfer with a pulse time $T=T_{\rm seq}$. Obtaining this value, however, indicates that one is (almost completely) unable to take advantage of wave propagation in order to speed up the state transfer and instead has to rely on the sequential neighbor-to-neighbor state transfer protocol.

\section{Conclusion}
\label{Sec:Conclusion}

We have considered the effect of disorder in a spin chain on the controlled transfer of an excitation or an unknown quantum state from one end of the chain to the opposite end of the chain, particularly in the case where the control fields can be applied only to the two sites at the two opposite ends of the chain. We have found that with optimized pulses one can achieve time-efficient perfect state transfer in the vast majority of disorder instances up to disorder strengths for which the uncontrolled dynamics would allow the excitation to propagate a distance that is only a small fraction of the total length of the chain. The application of quantum-repeater-like ideas where control fields are also applied to a number of intermediate sites enhances the transfer fidelity, allowing the protocol to give a high fidelity for rather large values of the disorder strength. However, the usefulness of this approach decreases when the disorder strength exceeds some value that depends on the details of the protocol.

These results demonstrate that local control can counter the localizing effects of disorder even substantially away from the points where the control fields are applied. Nevertheless, at sufficiently high disorder strengths the ability to perform state transfer deteriorates and even with full on-site control one cannot do much better than the neighbor-to-neighbor transfer protocol as the disorder strength approaches its maximum value.

We note here that we have only considered static disorder. In practice, dynamical disorder, or in other words temporal fluctuations of the parameters, are also present in many physical settings \cite{BurgarthStateTransferWithDisorder2}. While it is possible that the optimized pulses in the presence of such fluctuations will differ from those that are optimal in the absence of fluctuations, we believe that it is unlikely that control signals can reduce the effects of an unknown time-dependent noise signal.

In conclusion, our results provide theoretical understanding to the question of the competition between the constraining effects of disorder and the enabling effects of control. With the steady experimental progress in quantum coherence in spins and artificial atoms \cite{Ladd} and the high interest in constructing spin chains for fundamental investigation and quantum state transfer purposes, it is likely that such systems will be constructed in the near future, which will then allow the ideas discussed here to be tested and applied to achieve high-fidelity quantum state transfer in these systems.

We would like to thank Yu Chen for useful discussions that inspired a number of ideas presented in this work.


\begin{thebibliography}{99}

\bibitem{Anderson} P. W. Anderson, Phys. Rev. {\bf 109}, 1492 (1958).

\bibitem{AndersonLocalizationExperiments} J. Billy, V. Josse, Z. C. Zuo, A. Bernard, B. Hambrecht, P. Lugan, C. Clement, L. Sanchez-Palencia, P. Bouyer, and A. Aspect, Nature {\bf 453}, 891 (2008); G. Roati, C. D. Errico, L. Fallani, M. Fattori, C. Fort, M. Zaccanti, G. Modugno, M. Modugno, and M. Inguscio, Nature {\bf 453}, 895 (2008); S. Longhi, Laser \& Photon. Rev. {\bf 3}, 243 (2009); I. M. Georgescu, S. Ashhab, and F. Nori, Rev. Mod. Phys. {\bf 86}, 153 (2014).

\bibitem{Bose} S. Bose, Phys. Rev. Lett. {\bf 91}, 207901 (2003).

\bibitem{Christandl} M. Christandl, N. Datta, A. Ekert, and A. J. Landahl, Phys. Rev. Lett. {\bf 92}, 187902 (2004); see also L. Banchi, T. J. G. Apollaro, A. Cuccoli, R. Vaia, P. Verrucchi, Phys. Rev. A {\bf 82}, 052321 (2010).

\bibitem{BoseReview} S. Bose, Contemporary Physics {\bf 48}, 13 (2007); L. Banchi, Eur. Phys. J. Plus {\bf 128}, 137 (2013).

\bibitem{BurgarthStateTransferWithControl} D. Burgarth, V. Giovannetti, and S. Bose, Phys. Rev. A {\bf 75}, 062327 (2007).

\bibitem{QuantumRouters} A. D. Greentree, S. J. Devitt, L. C. L. Hollenberg, Phys. Rev. A {\bf 73}, 032319 (2006); E. Jonckheere, F. C. Langbein, S. Schirmer, Quantum Inf. Process. {\bf 13}, 1607 (2014).

\bibitem{BurgarthStateTransferWithDisorder1} One study included small amounts of disorder in order to demonstrate the robustness of a dual-rail protocol for perfect state transfer; D. Burgarth and S. Bose, New J. Phys. {\bf 7}, 135 (2005).

\bibitem{BurgarthStateTransferWithDisorder2} One study investigated state transfer in the presence of fluctuations that are static during a single run of the protocol but change on a slower timescale and therefore can vary from run to run; D. Burgarth, Eur. Phys. J. Special Topics {\bf 151}, 147 (2007).

\bibitem{Wu} See also L.-A. Wu, Y.-X. Liu, and F. Nori, Phys. Rev. A {\bf 80}, 042315 (2009); Z.-M. Wang, L.-A. Wu, M. Modugno, W. Yao, and B. Shao, Sci. Rep. {\bf 3}, 3128 (2013); D. de Falco, D. Tamascelli, J. Phys. A: Math. Theor. {\bf 46}, 225301
(2013).

\bibitem{Khaneja} N. Khaneja, T. Reiss, C. Kehlet, T. S. Herbr\"uggen, S. J. Glaser, J. Magn. Reson. {\bf 172}, 296 (2005).

\bibitem{PulseFootnote} Note that by ``control pulses'' we mean the control signals applied over the entire duration of the transfer protocol and not short pulses that are applied at specific points in time during the protocol.

\bibitem{Ashhab} S. Ashhab, P. C. de Groot, and F. Nori, Phys. Rev. A {\bf 85}, 052327 (2012).

\bibitem{GRAPEFootnote} The GRAPE algorithm searches (using a gradient-based approach starting from a randomly generated initial guess) for signals [e.g.~$\omega_1(t)$ and $\omega_N(t)$] that are piece-wise-constant functions of time, which in our case have 200 segments (and we have verified that our results remain essentially unchanged if we increase this number to 1000). Except for being piece-wise constant, no further constraints are imposed on the signals. In particular, the algorithm does not assume or require that the signals at the different sites [e.g.~$\omega_1(t)$ and $\omega_N(t)$] are correlated.

\bibitem{ComputationTimeFootnote} A single calculation with the parameters mentioned in the main text takes a computation time on the order of ten hours on a typical present-day single processor using the software package Mathematica.

\bibitem{BurgarthCouplingStrengthEstimation} D. Burgarth, K. Maruyama, and F. Nori, Phys. Rev. A {\bf 79}, 020305(R) (2009).

\bibitem{NonmonotonicProbabilityFootnote} The fact that in some cases the probability exhibits some non-monotonicity must be a computational artefact. Indeed, if a certain fidelity can be achieved for pulse time $T$, the same fidelity can be obtained for any longer time $T'$ by tuning the edge qubits far away from resonance with the rest of the chain for a total duration of $T'-T$ at the beginning or end of the pulse, thus dividing the total pulse time $T'$ into an active pulse time of duration $T$ and an idle time of duration $T'-T$. Instances of non-monotonicity were rare in our calculations, leading us to infer that numerical errors were probably at the level of a few percent in most cases. The notable exception is the case with a very small amount of disorder. We suspect that the large dip that appeared in all the simulations is related to the degree of matching between the natural wave propagation speed and the chain length. More specifically, when the pulse time is chosen to be close to the time that a maximum-speed wave needs to traverse the chain, a simple control pulse will give a high transfer probability, and the pulse search algorithm will find such pulses more easily than in the case of poor matching between the pulse time and chain length. We also note here that our calculations exhibited slower convergence in the case $\Delta=0.1$ than with higher values of $\Delta$.

\bibitem{IPR} R. J. Bell and P. Dean, Discuss. Faraday Soc. {\bf 50}, 55 (1970); W. M. Visscher, J. Non-Cryst. Sol. {\bf 8-10}, 477 (1972); see also J. Biddle, D. J. Priour, B. Wang, and S. Das Sarma, Phys. Rev. B {\bf 83}, 075105 (2011).

\bibitem{Ladd} see e.g. T. D. Ladd, F. Jelezko, Y. Nakamura, R. Laflamme, C. Monroe, and J. L. O'Brien, Nature {\bf 464}, 45 (2010); I. Buluta, S. Ashhab, and F. Nori, Rep. Prog. Phys. {\bf 74}, 104401 (2011); R. Barends, J. Kelly, A. Megrant, A. Veitia, D. Sank, E. Jeffrey, T. C. White, J. Mutus, A. G. Fowler, B. Campbell, Y. Chen, Z. Chen, B. Chiaro, A. Dunsworth, C. Neill, P. O'Malley, P. Roushan, A. Vainsencher, J. Wenner, A. N. Korotkov, A. N. Cleland, J. M. Martinis, Nature {\bf 508}, 500 (2014); P. Richerme, Z.-X. Gong, A. Lee, C. Senko, J. Smith, M. Foss-Feig, S. Michalakis, A. V. Gorshkov, and C. Monroe, Nature {\bf 511}, 198 (2014); P. Jurcevic, B. P. Lanyon, P. Hauke, C. Hempel, P. Zoller, R. Blatt, and C. F. Roos, Nature {\bf 511}, 202 (2014); S. Viciani, M. Lima, M. Bellini, and F. Caruso, Phys. Rev. Lett. {\bf 115}, 083601 (2015).

\end{thebibliography}
\end{document}